\RequirePackage[displaymath]{lineno}

\documentclass[twocolumn,aps,prl,showpacs,superscriptaddress,floatfix,nofootinbib]{revtex4}
\usepackage{newtxtext,newtxmath,booktabs,siunitx}
\usepackage{dcolumn}
\usepackage{bm}
\usepackage{float}
\usepackage[]{graphicx}  
\usepackage{booktabs}
\usepackage{tabularx}
\usepackage{amsmath}
\usepackage{xcolor}
\usepackage{multirow}
\usepackage[colorlinks,citecolor=blue,urlcolor=blue,linkcolor=blue]{hyperref}

\usepackage[displaymath]{lineno}

\newcommand {\snn}      {\sqrt{s_{_{\rm NN}}}}

\newcommand {\Nch}      {N_{\rm ch}}

\newcommand {\Zr}       {$^{96}$Zr}
\newcommand {\Ru}       {$^{96}$Ru}
\newcommand {\RuRu}     {$^{96}_{44}$Ru+$^{96}_{44}$Ru}
\newcommand {\ZrZr}     {$^{96}_{40}$Zr+$^{96}_{40}$Zr}
\newcommand {\Pb} {$^{208}$Pb}

\newcommand {\rnp}      {$\Delta r_{\rm np}$}

\newcommand {\hijing}   {Hijing}
\newcommand {\trento}   {{\sc trento}}
\newcommand {\rqmd}     {{\sc urqmd}}
\newcommand {\pythia}     {{\sc pythia}}

\newcommand {\rZr} {q_{\rm ZrZr}}
\newcommand {\rRu} {q_{\rm RuRu}}
\newcommand {\dQ} {\Delta Q}

\newcommand {\dBp} {\Delta B_{p}}

\begin{document}
\title{Measuring neutron skin by grazing isobaric collisions}
\author{Hao-jie Xu\footnote{%
        Corresponding author: haojiexu@zjhu.edu.cn}
 }
\affiliation{School of Science, Huzhou University, Huzhou, Zhejiang 313000, China}
\author{Hanlin Li\footnote{%
        Corresponding author: lihl@wust.edu.cn}
        }
\affiliation{College of Science, Wuhan University of Science and Technology, Wuhan, Hubei 430065, China}
\author{Ying Zhou}
\affiliation{
School of Physics and Astronomy, Shanghai Key Laboratory for Particle Physics and Cosmology, and Key Laboratory for Particle Astrophysics and Cosmology (MOE), Shanghai Jiao Tong University, Shanghai 200240, China
}
\author{Xiaobao Wang}
\affiliation{School of Science, Huzhou University, Huzhou, Zhejiang 313000, China}
\author{Jie Zhao}
\affiliation{Department of Physics and Astronomy, Purdue University, West Lafayette, Indiana 47907, USA}
\author{Lie-Wen Chen\footnote{%
        Corresponding author: lwchen@sjtu.edu.cn}
}
\affiliation{
School of Physics and Astronomy, Shanghai Key Laboratory for Particle Physics and Cosmology, and Key Laboratory for Particle Astrophysics and Cosmology (MOE), Shanghai Jiao Tong University, Shanghai 200240, China
}
\author{Fuqiang Wang\footnote{%
        Corresponding author: fqwang@purdue.edu}
}
\affiliation{School of Science, Huzhou University, Huzhou, Zhejiang 313000, China}
\affiliation{Department of Physics and Astronomy, Purdue University, West Lafayette, Indiana 47907, USA}

\begin{abstract}
Neutron skin thickness (\rnp) of nuclei and the inferred nuclear symmetry energy are of critical importance to nuclear physics and astrophysics. It is  traditionally measured by nuclear processes with significant theoretical uncertainties. We recently proposed an indirect measurement of the \rnp\ by charged hadron multiplicities in central isobaric collisions at relativistic energies, which are sensitive to nuclear densities. In this Letter, we propose a direct measurement of the \rnp\ by using net-charge multiplicities in ultra-peripheral (grazing) collisions of those isobars, under the assumption that they are simple superimposition of nucleon-nucleon interactions. We illustrate this novel approach by the \trento\ and \rqmd\ models. 
\end{abstract}

\maketitle

Nuclei are bound states of protons and neutrons by the overall attractive nuclear force. 
The nuclear strong force is isospin symmetric, however, because of Coulomb interactions, heavy nuclei usually need more neutrons than protons to remain stable.
The root-mean-square radius of neutron distribution in a heavy nucleus is thus larger than that of the proton distribution. The difference is referred to as the neutron skin thickness, $\Delta r_{\rm np}\equiv r_{n}-r_{p}$~\cite{Brown:2000pd}.
With more neutrons comes the penalty symmetry energy associated with the asymmetry between the proton and neutron numbers. 
By measuring the \rnp, one gains valuable information about the nuclear symmetry energy. 
Of particular interest is the symmetry energy density slope parameters $L$ at the nuclear saturation density $\rho_0$~\cite{Chen:2005ti,RocaMaza:2011pm,Tsang:2012se,Horowitz:2014bja} and $L_c$ at the critical density $\rho_c \approx 0.11$~fm$^{-3}$~\cite{Zhang:2013wna}.  

The \rnp\ has traditionally been measured by low-energy electron and hadron scatterings off nuclei~\cite{Frois:1987hk,Lapikas:1003zz,RocaMaza:2011pm,Tsang:2012se,Tarbert:2013jze}. Because of the inevitable uncertainties in modeling the strong interaction of the scattering processes in quantum chromodynamics (QCD)~\cite{Ray:1992fj}, large uncertainties on the $L$ and $L_c$ persist. 
Void of the strong interaction uncertainties, parity-violating scattering processes with electrons~\cite{Donnelly:1989qs,Horowitz:1999fk} and neutrinos~\cite{Akimov:2017ade}, sensitive to the neutral current, have been measured. The latest result from the Lead Radius Experiment (PREX-II) on the \Pb, \rnp\ = $0.283\pm0.071$~fm~\cite{Abrahamyan:2012gp,Adhikari:2021phr}, still has a large statistical uncertainty. This leads to $L_c=71.5\pm22.6$ and $L=105\pm37$~MeV\cite{Reed:2021nqk}, compared to  $L=75\pm25$~MeV\cite{Centelles:2008vu} from traditional scattering experiments, where the uncertainty is dominated by statistical one in the former and the systematic one in the latter. Efforts combining the PREX-II and astro-nuclear observations may improve the $L_c$ and $L$ constraints~\cite{Yue:2021yfx}.

Recently, we have proposed to use the ratio of charged hadron multiplicities ($\Nch$) in most central collisions of isobars (\RuRu\ and \ZrZr) at the Relativistic Heavy Ion Collider (RHIC) to help constrain the \rnp~\cite{Li:2019kkh}. This exploits the sensitivity of particle production to nucleon density distribution, which differs slightly between Ru and Zr because of their different \rnp\ values. 
Note that $\Nch$ is isospin insensitive; it is essentially the same among $pp$, $pn$, and $nn$ interactions at high energies (see Table~\ref{tab:pythia}). 
Our idea in Ref.~\cite{Li:2019kkh} probes the nucleon density and, in turn, the neutron's knowing the precisely measured proton's; it does not directly probe the neutron density.

The nucleon density difference is, of course, also present in peripheral collisions but
there, the $\Nch$ sensitivity to \rnp\ is weak (see Fig.~4 of Ref.~\cite{Li:2018oec} where the isobar difference is only 2\%).
However, those ultra-peripheral collisions, where the nuclei are only grazing each other, must have very different mixture of participant protons and neutrons, and therefore likely yield significantly different net-charge numbers ($\dQ$). 
This is obvious especially for the case of full acceptance with exact charge conservation, in which a proton releases one net-charge number, whereas the contribution from neutrons is none.
This difference should also manifest in limited-acceptance midrapidity detectors. By measuring $\dQ$ in those grazing collisions, one could in principle determine the difference in the numbers of protons and neutrons participating in those collisions, which is {\em directly} sensitive to \rnp. 

To make our idea more concrete, we list in Table~\ref{tab:pythia} the
$\dQ$ (and $\Nch$ as well as the net-proton number $\dBp$) in minimum bias $pp$, $pn$, and $nn$ interactions at $\sqrt{s}=200$~GeV, simulated by the \pythia\ event generator~\cite{Sjostrand:2006za,Sjostrand:2007gs} (version 8.240). We used Monash tune~\cite{Skands:2014pea} which was based on the LHC data (we will discuss this particular point at end of the paper).
The acceptance cuts for $\dQ$ are $|\eta|<1$ and $0.2 < p_T < 2$ GeV/$c$, and we have excluded the (anti-)protons with $p_T<0.4$ GeV/$c$ (where the protons are strongly contaminated by background in experiments~\cite{Abelev:2008ab}). 
The $\Nch$ is from $|\eta|<0.5$ and $0.2 < p_T < 2$ GeV/$c$ (as usually performed in experiment). 
The $\dQ$ differs significantly between $pp$ and $nn$ interactions, whereas $\dBp$ is almost the same. The charge difference in the initial baryons that have transported to midrapidity is transferred
nearly entirely to mesons. 
This difference will imprint in peripheral nucleus-nucleus (AA) collisions, the magnitude of which depends on \rnp.
If a peripheral AA collision is a simple superimposition of nucleon-nucleon (NN) interactions, then one can predict $\dQ$ in AA by those in NN collisions as
\begin{equation}
    \dQ_{AA}\propto q_{AA}\dQ_{pp} + (1-q_{AA})\dQ_{nn}\,,
    \label{eq:dQ}
\end{equation}
where $q_{AA}$ is the fraction of protons among the participant nucleons.
The proportionality (normalization) factor is simply the number of NN collisions.
We have taken $\dQ_{pn}=(\dQ_{pp}+\dQ_{nn})/2$ which is a good assumption as seen in Table~\ref{tab:pythia}-- 
we have also verified this in the \hijing\ model~\cite{Wang:1991hta,Wang:1996yf} and the \rqmd\ model~\cite{Bass:1998ca,Bleicher:1999xi}. One can have a similar equation for $\dBp$, but since $\dBp$'s are almost the same among $pp$, $pn$ and $nn$ interactions it is of little use for our purpose.

\begin{table}
      \caption{The midrapidity charged hardon multiplicity $\Nch$ ($|\eta|<0.5$), the net-charge $\dQ$, and net-proton $\dBp$ within $|\eta|<1$ and $0.2<p_T<2$~GeV/$c$ (excluding protons and antiprotons from $p_T<0.4$~GeV/$c$) in minimum bias $pp$, $pn$, and $nn$ interactions at $\sqrt{s}=200$~GeV by \pythia\ (version 8.240, Monash tune~\cite{Skands:2014pea}).\label{tab:pythia}}
\begin{centering}
\begin{tabular}{c|ccc}
\hline
    \pythia &  $\Nch$ & $\dQ$  &  $\dBp$  \\ 
\hline
p+p &  $3.487$ & $0.0793$  &  $0.0158$ \\ 
\hline
    p+n & $3.485$ & $0.0255$   &  $0.0144$  \\ 
\hline
n+n &  $3.483$ & $-0.0279$  &  $0.0130$ \\ 

\hline
\end{tabular}
\par\end{centering}
\end{table}

Now consider the isobar collisions of Ru+Ru and Zr+Zr at $\snn=200$~GeV. 
The $\dQ$ ratio in Ru+Ru over Zr+Zr collisions, under the superimposition assumption, is
\begin{equation}
      R_{\dQ}\equiv\frac{\dQ_{\rm RuRu}}{\dQ_{\rm ZrZr}} 
      = \frac{\rRu + \alpha/(1-\alpha)}{\rZr + \alpha/(1-\alpha)}\,, 
      \label{eq:R}
\end{equation}
where $\alpha \equiv \Delta Q_{nn}/\Delta Q_{pp}$ is the $\dQ$ ratio in $nn$ to $pp$ interactions; \pythia\ gives $\alpha\simeq -0.352$.
The overall $\rRu$ and $\rZr$ values for the whole nuclei are $44/96$ and $40/96$, respectively;
they would give $R_{\dQ}\simeq 1.267$. Of course, the simple superimposition assumption breaks down in non-peripheral collisions because of nuclear effects. However, the assumption should be good for grazing AA collisions, where one expects Eq.~(\ref{eq:R}) to be valid.
The general idea to probe \rnp\ by $R_{\dQ}$ is that
a sizable \rnp\ will make the $q_{AA}$ decrease dramatically with increasing impact parameter ($b$) in those grazing collisions. 
\rnp\ of \Zr\ is significantly larger than that of \Ru, so the $R_{\dQ}$ ratio amplifies the \rnp\ sensitivity. The \rnp\ of both nuclei are controlled by the $L_c$ parameter, thus a measurement of $R_{\dQ}$ can determine its value.

Following our previous work~\cite{Li:2019kkh,Xu:2017zcn}, we examine four sets of \Ru\ and \Zr\ nuclear densities from energy density functional theory (DFT). One is the standard Skyrme-Hartree-Fock (SHF) model (see, e.g., Ref.~\cite{Chabanat:1997qh}) using the well-known interaction set SLy4~\cite{Chabanat:1997un,Wang:2016rqh}.
The other is the extended SHF (eSHF) model~\cite{Chamel:2009yx,Zhang:2015vaa} with three sets of interaction parameters, denoted as Lc47, Lc20 and Lc70, corresponding to $L_c= 47.3$, 20 and 70~MeV~\cite{Zhang:2014yfa}, respectively, all with the symmetry energy $E_{\rm sym}(\rho_c) = 26.65$~MeV~\cite{Zhang:2013wna}. The Lc47 set is the best fit to data on the nuclear masses and electric dipole polarizibility in $^{208}$Pb~\cite{Zhang:2013wna}, and the other two are to explore the effects of the symmetry energy (and neutron skin) variations.

\begin{figure} [!htb]
      \includegraphics[scale=0.42]{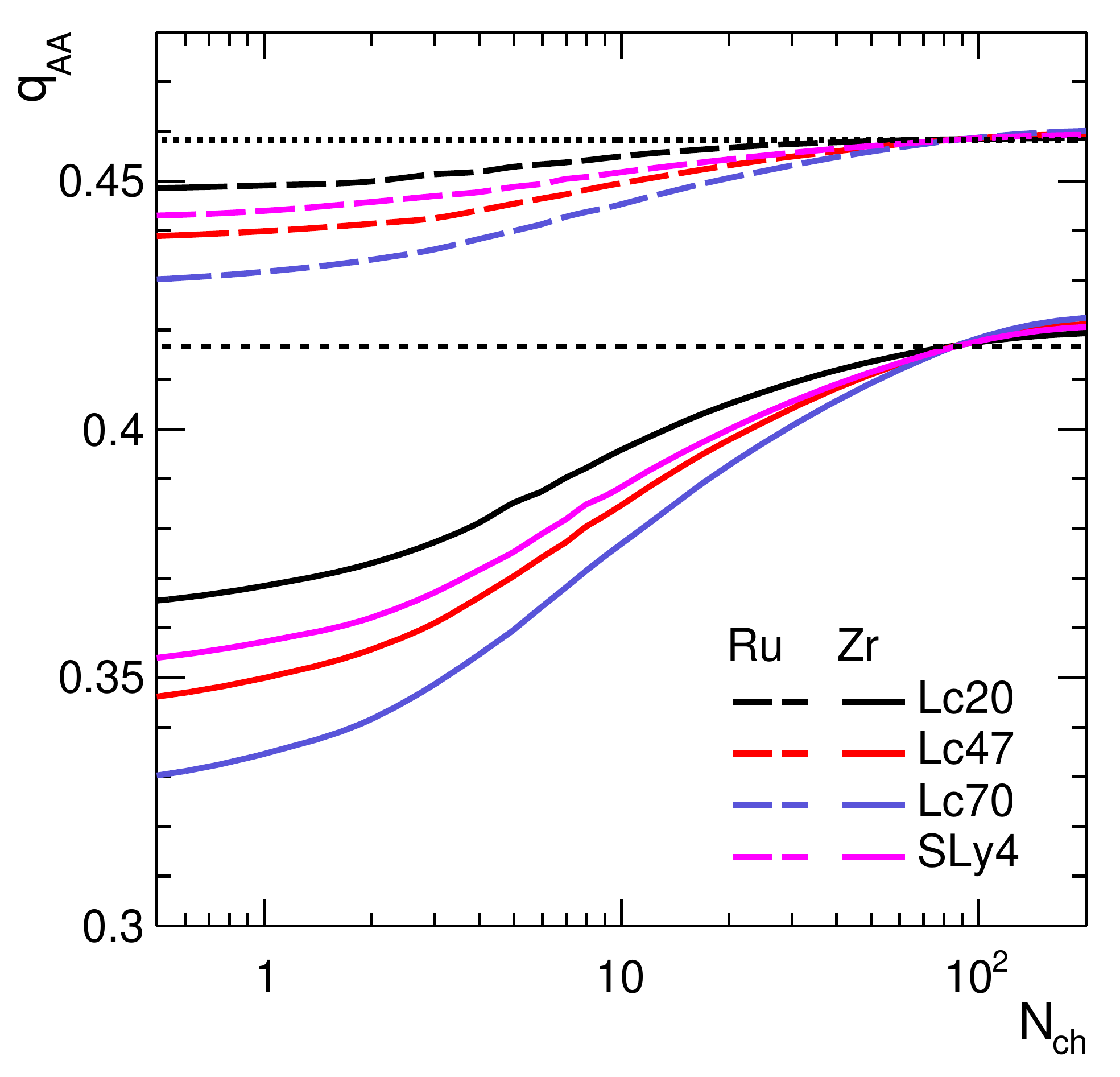}
      \caption{(Color online).
      The proton fractions $q_{AA}$ among participant nucleons as function of charged hadron multiplicity $\Nch$ calculated by \trento\  with the \Ru\ and \Zr\ nuclear densities from eSHF (Lc20, Lc47, Lc70) and SHF (SLy4). The dotted and dashed lines indicate the overall values of 44/96 and 40/96 of the entire Ru and Zr nuclei, respectively.
      \label{fig:rxx}} 
 \end{figure}

With a given nuclear density, one can calculate the $q_{AA}$ parameter in Ru+Ru and Zr+Zr collisions as a function of $b$ (and $\Nch$). We use the \trento\ model~\cite{Moreland:2014oya} to do that. In \trento\ particle production is related to the reduced thickness, $\Nch\propto T_{R}(p;T_{A},T_{B})\equiv[(T_{A}^{p} + T_{B}^{p})/2]^{1/p}$~\cite{Bernhard:2016tnd,Moreland:2014oya}.
We use the parameter $p=0$ (i.e., $\Nch\propto\sqrt{T_AT_B}$), a gamma fluctuation parameter $k=1.4$,
and a Gaussian nucleon size of $0.6$ fm, which were found to well describe the multiplicity data in heavy ion collisions~\cite{Bernhard:2016tnd,Moreland:2014oya}. 
The $\rRu$ and $\rZr$ from \trento\ calculations are shown in Fig.~\ref{fig:rxx}. 
They are sensitive to \rnp\ and thus the $L_c$ parameter, and decrease with decreasing $\Nch$ in peripheral collisions. Both these features indicate the effect of \rnp\ on $R_{\dQ}$. The sensitivity is larger in Zr+Zr than Ru+Ru collisions because of the larger \rnp\ of Zr.

\begin{figure} [!htb]
      \includegraphics[scale=0.42]{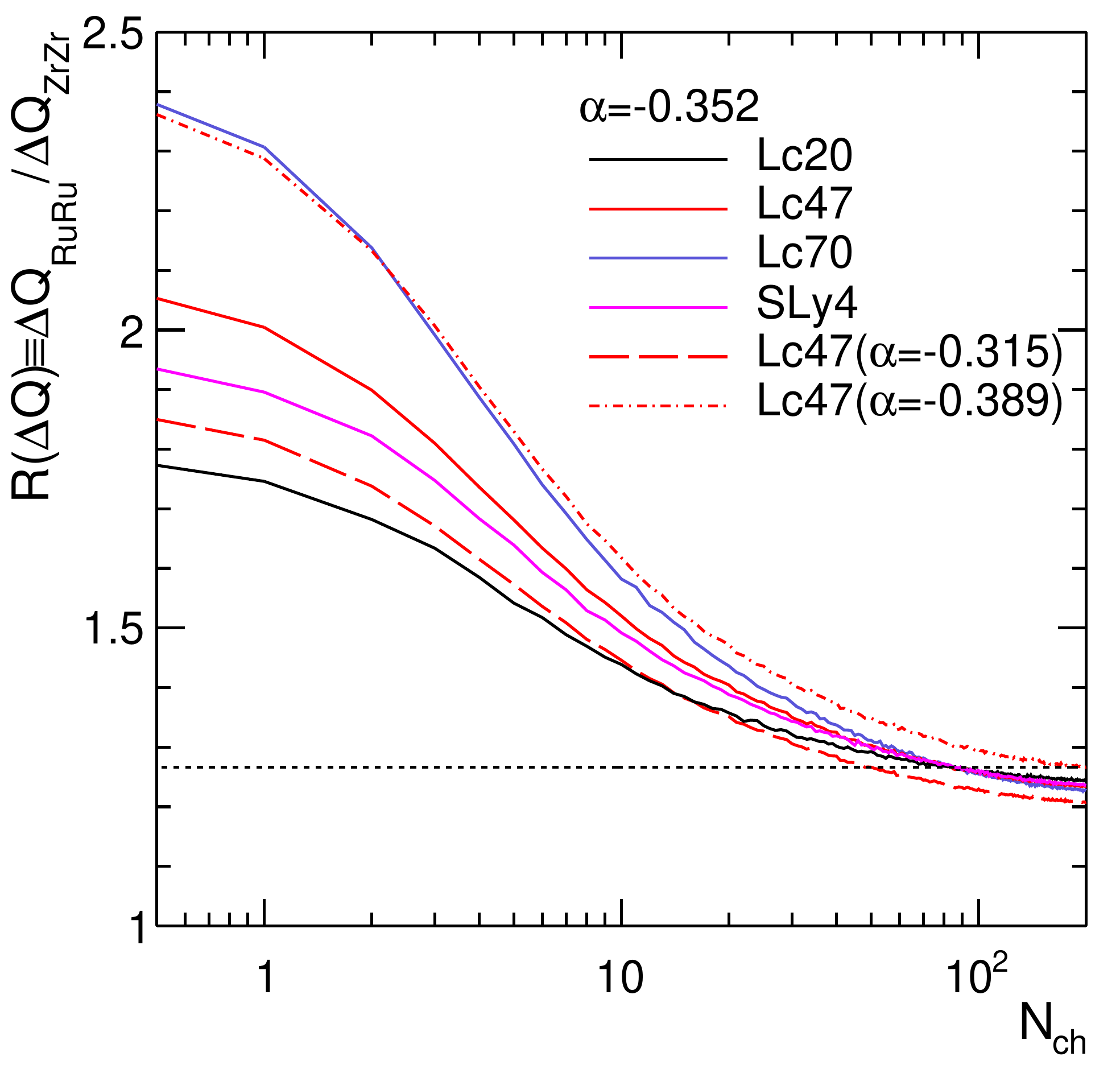}
      \caption{(Color online).
	$R_{\dQ}$ as function of $\Nch$ by Eq.~(\ref{eq:R}) with $\alpha=-0.352$ from \pythia\ and the $q_{AA}$ from \trento\ in Fig.~\ref{fig:rxx} with four DFT densities (solid curves). The dashed (dash-dotted) curve uses $\alpha=-0.315$ ($-0.389$) with the Lc47 density. The dotted line indicates the overall value of 1.267.
      \label{fig:RQTrento}} 
\end{figure}

With those ratios and the $\alpha$ value, the $R_{\dQ}$ can be obtained by Eq.~(\ref{eq:R}) directly. Figure~\ref{fig:RQTrento} shows the $R_{\dQ}$ as a function of $\Nch$ using $\alpha=-0.352$ from \pythia\ (Table~\ref{tab:pythia}, the dashed curve using another $\alpha$ value will be discussed later). 
The differences among the nuclear densities are obvious at small $\Nch$.
Although the calculated $R_{\dQ}$ is unlikely correct at large $\Nch$ because of the invalid assumption of simple NN superimposition for those non-peripheral AA collisions, the $R_{\dQ}$ for grazing collisions with small $\Nch$ should be robust and can be used to determine the \rnp\ and thus the $L_c$ parameter.
We note that the $\alpha$ value is obtained from minimum bias NN interactions. However, a grazing AA collision with a particular $\Nch$ value selects biased underlying NN interactions, especially when $\Nch$ is small. We have verified that $\dQ_{pp}$ and $\dQ_{nn}$ do indeed depend on $\Nch$ of NN collisions, however, their ratio $\alpha$ is insensitive to $\Nch$. Using a constant $\alpha$ in Eq.~(\ref{eq:R}) is, therefore, justified.

\begin{figure} [!htb]
      \includegraphics[scale=0.42]{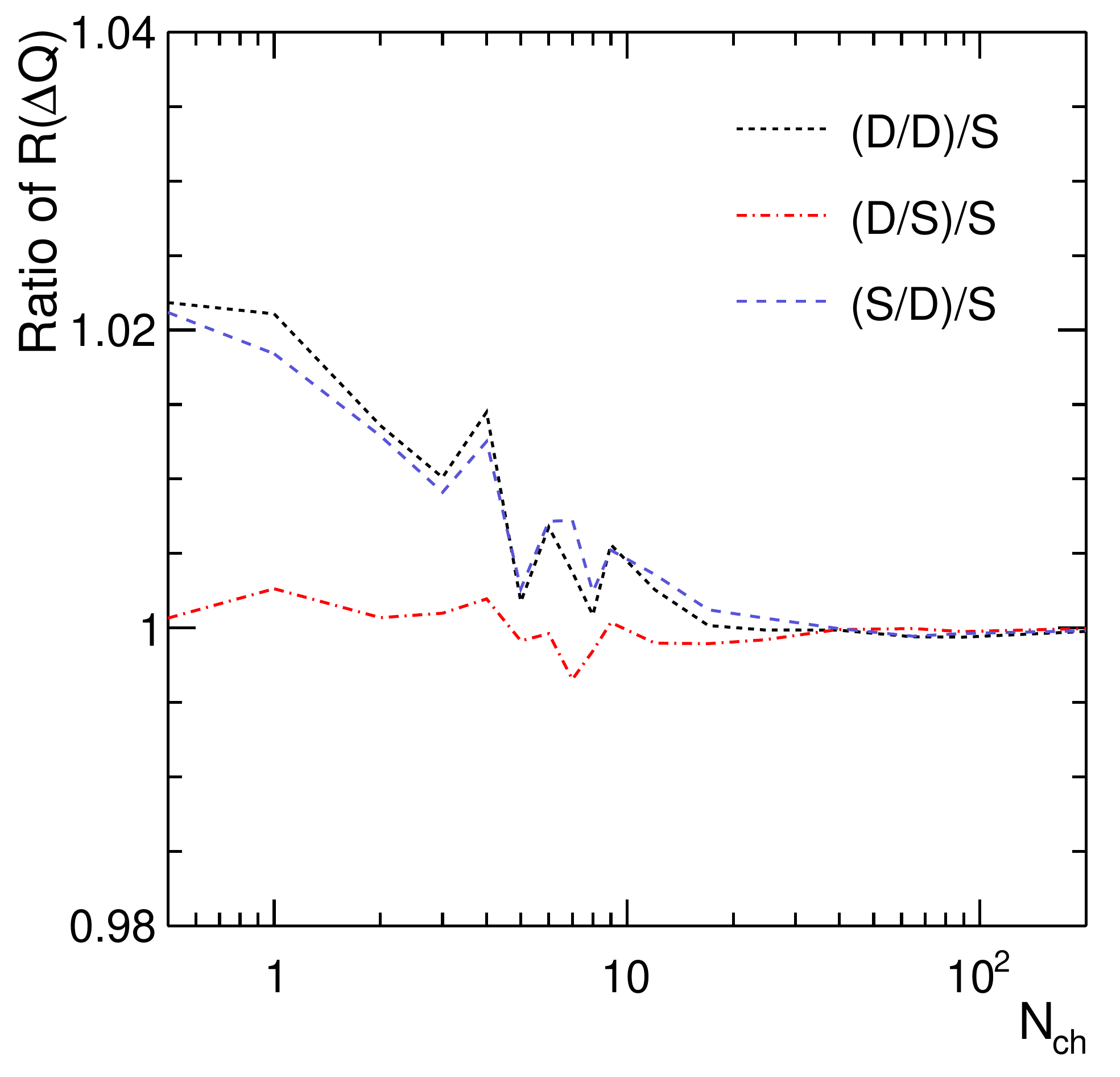}
      \caption{(Color online).
      Effect of nuclear quadrupole deformity on the $R_{\dQ}$, shown by its relative deviation from the spherical case. 
      \label{fig:RQTrentoDeformty}} 
\end{figure}

The DFT densities we used have assumed spherical nuclei without considering the possible deformation in their calculations. 
Their experimental measurements of the quadrupole deformation parameter are also uncertain, ranging from $\beta_2=0.053$ to $0.158$ for Ru and from $0.08$ to $0.217$ for Zr~\cite{Deng:2016knn}. To test the sensitivity of our results to nuclear deformation, we follow Ref.~\cite{Xu:2021vpn} using the Woods-Saxon parametrization 
$\rho(r,\theta)\propto[1+\exp([r-R(1+\beta_{2}Y_{2}^{0}(\theta))]/a)]^{-1}$ with $\beta_2=0.16$ fixed and the $R$ and $a$ reproducing the first and second radial moments of the DFT-calculated spherical proton and neutron densities of SHF/SLy4. 
The changes in $R_{\dQ}$ from the spherical case are shown in Fig.~\ref{fig:RQTrentoDeformty} for the three combinations of deformed/spherical Ru and Zr nuclei. The effect is only a couple of percent, leading possibly to less than 5 MeV uncertainty in the extracted $L_c$ parameter. This small effect is not a surprise because the $R_{\dQ}$ is sensitive only to the relative proton/neutron composition on the nuclear surface, which is not sensitive to nuclear deformation. 

\begin{figure} [!htb]
      \includegraphics[scale=0.42]{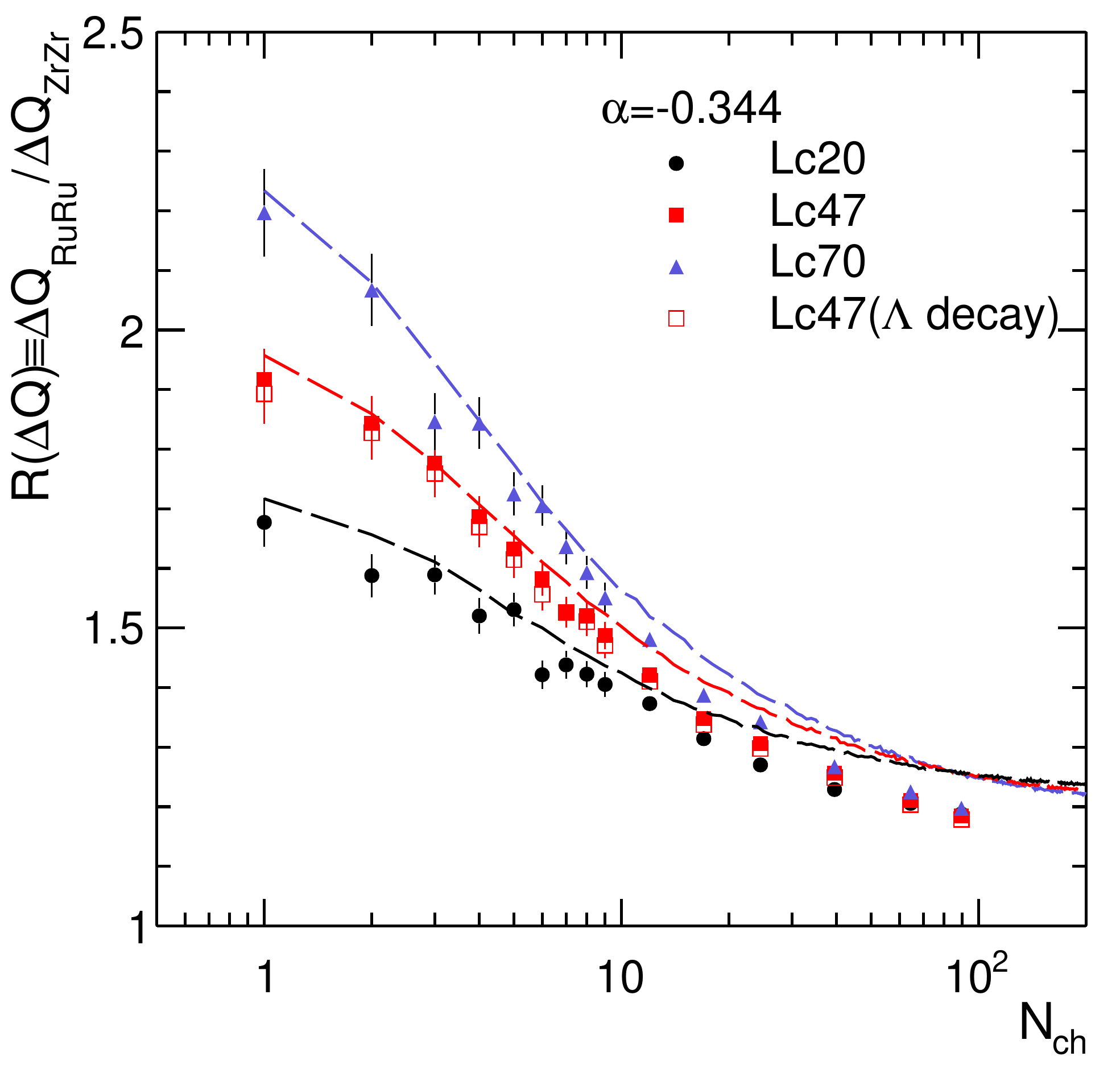}
      \caption{(Color online).
$R_{\dQ}$ for $b\in[7,20]$fm as function of $\Nch$ ($|\eta|<0.5$) simulated by \rqmd\ with DFT nuclear densities from eSHF (Lc20, Lc47, Lc70). $\dQ$ is calculated from $|\eta|<1$ and $0.2<p_{T}<2$ GeV/$c$ excluding the (anti-)protons with $p_{T}<0.4$ GeV/$c$.
The open red squares show a calculation for the Lc47 case including $\Lambda$-hyperon decays.
The curves are Eq.~(\ref{eq:R}) with $\alpha=-0.344$ from \rqmd\ NN interactions and the $q_{AA}$ from \trento\ in Fig.~\ref{fig:rxx}.}
      \label{fig:RQUrQMD}
\end{figure}

Next we examine our idea using a dynamical model.
We use the \rqmd\ (Ultra relativistic Quantum Molecular Dynamics, v3.4)~\cite{Bass:1998ca,Bleicher:1999xi}
as it have been widely used to study the conserved charge number and its fluctuations in heavy ion collisions~\cite{Adamczyk:2013dal,Xu:2016qjd}.
We simulate events within $b\in[7,20]$~fm since we focus only on peripheral collisions.
The same acceptance cuts have been applied as performed in \pythia\ simulations. 
Figure~\ref{fig:RQUrQMD} shows $R_{\dQ}$ as a function of $\Nch$. Similar splittings are found at low $\Nch$ as in Fig.~\ref{fig:RQTrento}. 
\rqmd\ simulations of NN interactions indicate $\alpha=-0.344$. 
Using this $\alpha$ value, the predicted curves by Eq.~(\ref{eq:R}) are superimposed in Fig.~\ref{fig:RQUrQMD}. The curves can fairly well describe the \rqmd\ data.
This indicates that the grazing collisions in \rqmd\ with $\Nch\lesssim 10$ are indeed simple superimposition of NN interactions. This is not surprising as only a few nucleons participate in such a grazing AA collision so any nuclear effect would be negligible.
At higher $\Nch$ the \rqmd\ data points deviate from the curves, presumably because those collisions are not simple NN superimpositions any more.
It may also be viewed as that the effective $\alpha$ in central AA collisions, because of nuclear effects, is very different from the one calculated using single NN interactions.

The large differences among the nuclear densities are mostly due
 to the negative value of $\alpha$ as mentioned before.
We note that a larger $p_{T}$ cut on (anti)proton makes $\alpha$ more negative, e.g., $\alpha=-0.416$ with $p_{T}>0.8 {\rm GeV}/c$ on (anti)proton, which may make the $R_{\dQ}$ more
sensitive to \rnp.
However, excluding any ``net-charge''  will weaken the correlation between 
the initial protons from the incoming nuclei and the final-state net-charge observable, 
which would introduce stronger model dependence.
We also note that the finite negative $\alpha$ value is not merely because 
we exclude the (anti)proton with $p_{T}<0.4 {\rm GeV}/c$; we have checked that without any $p_T$ cut, the $\alpha$ value is $-0.370$ at mid-rapidity ($|\eta|<1.0$).
In our \rqmd\ simulations, the hyperon decays are not included. 
The hyperon decays can introduce net-charge because of the combined effect of the finite acceptance (which does not always cover both daughter particles from the decay) and the asymmetry of those decays.
We have calculated the effect of $\Lambda$ decays for the Lc47 case, and found that the effect is negligible (see the open red squares in Fig.\ref{fig:RQUrQMD}). 

We have simulated $\sim4.5\times10^{8}$ \rqmd\ events for each isobar system in Fig.~\ref{fig:RQUrQMD}. They give a statistical uncertainty on $L_c$ of roughly 9~MeV using data points at $\Nch\lesssim10$. 
It should be noted that our proposal requires measurements of ultra-peripheral collisions which are often difficult because of trigger inefficiencies in experiment, increasing in severity with decreasing $\Nch$. 
Over $2\times10^{9}$ minimum bias events have been collected at the RHIC for each isobar system; they would likely give similar statistics to our simulated \rqmd\ sample, given the typical experimental trigger inefficiency. Thus, the statistical uncertainty on the extracted $L_c$ using isobar data is expected to be $\sim$10~MeV.
Since our $R_{\dQ}$ observable is a ratio of two isobar systems, many experimental systematic effects cancel and would unlikely contribute a significant part to the overall experimental uncertainty.
We note that the electromagnetic interactions could affect the proposed method. However, in experiment, one can avoid electromagnetic interactions by triggering on minimum bias hadronic interactions only.

Our idea relies on the state of the art DFT calculation and the  assumption of simple NN superimposition for grazing isobar collisions. 
The theorectical uncertainty from the former may be gauged by the two parameter sets of SLy4 and Lc47 shown in Fig.~\ref{fig:RQTrento}, corresponding to similar $L_c$ values. Their difference would give an uncertainty of $\sim$6~MeV. This is rather modest compared to the  substantial uncertainties from QCD modeling of traditional scattering observables. 
Another theoretical uncertainty on density distributions comes from nuclear deformation; this is insignificant as indicated by Fig.~\ref{fig:RQTrentoDeformty}.

The assumption of simple NN superimposition  should be rather robust for grazing heavy ion collisions over a wide range of  relativistic energies. We have focused on $\snn=200$~GeV in this Letter because the isobar collision data recently taken at RHIC were at this energy~\cite{Skokov:2016yrj}.
We note that isospin-sensitive observables, such as the $\pi^+/\pi^-$ ratio, have been proposed long ago to probe high-density symmetry energy in low energy nuclear collisions (not particularly in grazing collisions)~\cite{Li:2002qx}.  
In our study of grazing AA collisions, the NN information imprint in $\dQ$ is carried over by the $\alpha$ parameter.
The $\alpha$ value we used in this Letter comes from a particular tune of \pythia\ and \rqmd. It is used only to illustrate our proposed method; its value should be taken with caution. For example, \pythia6 (version 6.416) tuned to the RHIC data~\cite{Zhang:2018hcp} would give $\alpha=-0.315$,
the result of which with the Lc47 density is shown by the dashed curve in Fig.~\ref{fig:RQTrento}.
\hijing\ gives $\alpha=-0.389$ which is shown by the dashed-dot curve in Fig.~\ref{fig:RQTrento} for Lc47.
This would result in an uncertainty on $L_c$ comparable to the range of the parameter sets in Fig.~\ref{fig:RQTrento} (which was constrained from traditional scattering measurements). 
The determination of the $\alpha$ parameter is therefore essential.
The $\dQ$ data in $pp$ interactions are abundant, that in $nn$ is experimentally not available. Although $nn$ interactions are hard to perform, their information may be indirectly accessed by proton-deuteron and/or deuteron-deuteron collisions, yet to be conducted.
In addition, one may gain valuable information on the $\alpha$ parameter from existing p-Au and d-Au data at RHIC.
Provided that the $\alpha$ parameter can be determined relatively precisely, our proposal will likely result in a $L_c$ value with an overall uncertainty of 10-15~MeV. This would be a substantial complement to traditional methods with the present overall $\sim$20~MeV uncertainty~\cite{Reed:2021nqk}. 

To summarize, 
we demonstrate that the net-charge ratio $R_{\dQ}$ in isobar \RuRu\ over \ZrZr\ collisions can be predicted from those in $pp$ and $nn$ interactions under the assumption that nucleus-nucleus AA collisions are a simple superimposition of NN interactions. We show using the \rqmd\ model that this assumption is valid for ultra-peripheral (grazing) collisions with $\Nch\lesssim 10$ within the midrapidity range of $|\eta|<0.5$. The predicted $R_{\dQ}$ depends on the proton and neutron densities, particularly on the neutron skin thickness \rnp. The $R_{\dQ}$ measurement in grazing isobar collisions, together with DFT calculations of nuclear densities can, therefore, determine the \rnp\ and, in turn, the symmetry energy density slope parameter $L_c$.

We have previously proposed~\cite{Li:2019kkh} an indirect observable, the $\Nch$ ratio in central isobar collisions, to probe the \rnp. This observable is isospin insensitive and is based on the fact that particle production in relativistic heavy ion collisions depends on the total nuclear (proton+neutron) density which has some sensitivity to the \rnp. 
The $R_{\dQ}$ observable proposed here is isospin sensitive, directly related to the \rnp. It is based on the simple superimposition of NN interactions for those grazing AA collisions, and depends on the difference (proton $-$ neutron) density which is highly sensitive to the \rnp.
Combining these two observables will yield stringent constrains on the \rnp\ and, thus, the $L_c$ parameter.

\section*{Acknowledgments}
HX thanks Dr.~Qiang Zhao and Dr.~Wenbin Zhao for useful discussions. FW thanks Dr.~Torbjörn Sjöstrand and  Dr.~Christian Bierlich for helpful information on Pythia. This work is supported in part by the National Natural Science Foundation of China (Grant Nos.~11905059, 11947410, 12035006, 12047568, 12075085, 11625521, U1732138, 11605054, 11505056), the Ministry of Science and Technology of China (Grant No.~2020YFE0202001),
National SKA Program of China No.~2020SKA0120300,
the Zhejiang Provincial Natural Science Foundation of China (Grant No.~LY21A050001), 
the Natural Science Foundation of Hubei Province (Grant No. 2019CFB563),
and the U.S.~Department of Energy (Grant No.~DE-SC0012910).   

\bibliography{ref}

\begin{thebibliography}{44}
\expandafter\ifx\csname natexlab\endcsname\relax\def\natexlab#1{#1}\fi
\expandafter\ifx\csname bibnamefont\endcsname\relax
  \def\bibnamefont#1{#1}\fi
\expandafter\ifx\csname bibfnamefont\endcsname\relax
  \def\bibfnamefont#1{#1}\fi
\expandafter\ifx\csname citenamefont\endcsname\relax
  \def\citenamefont#1{#1}\fi
\expandafter\ifx\csname url\endcsname\relax
  \def\url#1{\texttt{#1}}\fi
\expandafter\ifx\csname urlprefix\endcsname\relax\def\urlprefix{URL }\fi
\providecommand{\bibinfo}[2]{#2}
\providecommand{\eprint}[2][]{\url{#2}}

\bibitem[{\citenamefont{Brown}(2000)}]{Brown:2000pd}
\bibinfo{author}{\bibfnamefont{B.~A.} \bibnamefont{Brown}},
  \bibinfo{journal}{Phys. Rev. Lett.} \textbf{\bibinfo{volume}{85}},
  \bibinfo{pages}{5296} (\bibinfo{year}{2000}).

\bibitem[{\citenamefont{Chen et~al.}(2005)\citenamefont{Chen, Ko, and
  Li}}]{Chen:2005ti}
\bibinfo{author}{\bibfnamefont{L.-W.} \bibnamefont{Chen}},
  \bibinfo{author}{\bibfnamefont{C.~M.} \bibnamefont{Ko}}, \bibnamefont{and}
  \bibinfo{author}{\bibfnamefont{B.-A.} \bibnamefont{Li}},
  \bibinfo{journal}{Phys. Rev.} \textbf{\bibinfo{volume}{C72}},
  \bibinfo{pages}{064309} (\bibinfo{year}{2005}), \eprint{nucl-th/0509009}.

\bibitem[{\citenamefont{Roca-Maza et~al.}(2011)\citenamefont{Roca-Maza,
  Centelles, Vinas, and Warda}}]{RocaMaza:2011pm}
\bibinfo{author}{\bibfnamefont{X.}~\bibnamefont{Roca-Maza}},
  \bibinfo{author}{\bibfnamefont{M.}~\bibnamefont{Centelles}},
  \bibinfo{author}{\bibfnamefont{X.}~\bibnamefont{Vinas}}, \bibnamefont{and}
  \bibinfo{author}{\bibfnamefont{M.}~\bibnamefont{Warda}},
  \bibinfo{journal}{Phys. Rev. Lett.} \textbf{\bibinfo{volume}{106}},
  \bibinfo{pages}{252501} (\bibinfo{year}{2011}), \eprint{1103.1762}.

\bibitem[{\citenamefont{Tsang et~al.}(2012)}]{Tsang:2012se}
\bibinfo{author}{\bibfnamefont{M.~B.} \bibnamefont{Tsang}}
  \bibnamefont{et~al.}, \bibinfo{journal}{Phys. Rev.}
  \textbf{\bibinfo{volume}{C86}}, \bibinfo{pages}{015803}
  (\bibinfo{year}{2012}), \eprint{1204.0466}.

\bibitem[{\citenamefont{Horowitz et~al.}(2014)\citenamefont{Horowitz, Brown,
  Kim, Lynch, Michaels, Ono, Piekarewicz, Tsang, and
  Wolter}}]{Horowitz:2014bja}
\bibinfo{author}{\bibfnamefont{C.~J.} \bibnamefont{Horowitz}},
  \bibinfo{author}{\bibfnamefont{E.~F.} \bibnamefont{Brown}},
  \bibinfo{author}{\bibfnamefont{Y.}~\bibnamefont{Kim}},
  \bibinfo{author}{\bibfnamefont{W.~G.} \bibnamefont{Lynch}},
  \bibinfo{author}{\bibfnamefont{R.}~\bibnamefont{Michaels}},
  \bibinfo{author}{\bibfnamefont{A.}~\bibnamefont{Ono}},
  \bibinfo{author}{\bibfnamefont{J.}~\bibnamefont{Piekarewicz}},
  \bibinfo{author}{\bibfnamefont{M.~B.} \bibnamefont{Tsang}}, \bibnamefont{and}
  \bibinfo{author}{\bibfnamefont{H.~H.} \bibnamefont{Wolter}},
  \bibinfo{journal}{J. Phys.} \textbf{\bibinfo{volume}{G41}},
  \bibinfo{pages}{093001} (\bibinfo{year}{2014}), \eprint{1401.5839}.

\bibitem[{\citenamefont{Zhang and Chen}(2013)}]{Zhang:2013wna}
\bibinfo{author}{\bibfnamefont{Z.}~\bibnamefont{Zhang}} \bibnamefont{and}
  \bibinfo{author}{\bibfnamefont{L.-W.} \bibnamefont{Chen}},
  \bibinfo{journal}{Phys. Lett.} \textbf{\bibinfo{volume}{B726}},
  \bibinfo{pages}{234} (\bibinfo{year}{2013}), \eprint{1302.5327}.

\bibitem[{\citenamefont{Frois and Papanicolas}(1987)}]{Frois:1987hk}
\bibinfo{author}{\bibfnamefont{B.}~\bibnamefont{Frois}} \bibnamefont{and}
  \bibinfo{author}{\bibfnamefont{C.~N.} \bibnamefont{Papanicolas}},
  \bibinfo{journal}{Ann. Rev. Nucl. Part. Sci.} \textbf{\bibinfo{volume}{37}},
  \bibinfo{pages}{133} (\bibinfo{year}{1987}).

\bibitem[{\citenamefont{Lapikas}(1993)}]{Lapikas:1003zz}
\bibinfo{author}{\bibfnamefont{L.}~\bibnamefont{Lapikas}},
  \bibinfo{journal}{Nucl. Phys.} \textbf{\bibinfo{volume}{A553}},
  \bibinfo{pages}{297c} (\bibinfo{year}{1993}).

\bibitem[{\citenamefont{Tarbert et~al.}(2014)}]{Tarbert:2013jze}
\bibinfo{author}{\bibfnamefont{C.~M.} \bibnamefont{Tarbert}}
  \bibnamefont{et~al.}, \bibinfo{journal}{Phys. Rev. Lett.}
  \textbf{\bibinfo{volume}{112}}, \bibinfo{pages}{242502}
  (\bibinfo{year}{2014}), \eprint{1311.0168}.

\bibitem[{\citenamefont{Ray et~al.}(1992)\citenamefont{Ray, Hoffmann, and
  Coker}}]{Ray:1992fj}
\bibinfo{author}{\bibfnamefont{L.}~\bibnamefont{Ray}},
  \bibinfo{author}{\bibfnamefont{G.~W.} \bibnamefont{Hoffmann}},
  \bibnamefont{and} \bibinfo{author}{\bibfnamefont{W.~R.} \bibnamefont{Coker}},
  \bibinfo{journal}{Phys. Rept.} \textbf{\bibinfo{volume}{212}},
  \bibinfo{pages}{223} (\bibinfo{year}{1992}).

\bibitem[{\citenamefont{Donnelly et~al.}(1989)\citenamefont{Donnelly, Dubach,
  and Sick}}]{Donnelly:1989qs}
\bibinfo{author}{\bibfnamefont{T.~W.} \bibnamefont{Donnelly}},
  \bibinfo{author}{\bibfnamefont{J.}~\bibnamefont{Dubach}}, \bibnamefont{and}
  \bibinfo{author}{\bibfnamefont{I.}~\bibnamefont{Sick}},
  \bibinfo{journal}{Nucl. Phys.} \textbf{\bibinfo{volume}{A503}},
  \bibinfo{pages}{589} (\bibinfo{year}{1989}).

\bibitem[{\citenamefont{Horowitz et~al.}(2001)\citenamefont{Horowitz, Pollock,
  Souder, and Michaels}}]{Horowitz:1999fk}
\bibinfo{author}{\bibfnamefont{C.~J.} \bibnamefont{Horowitz}},
  \bibinfo{author}{\bibfnamefont{S.~J.} \bibnamefont{Pollock}},
  \bibinfo{author}{\bibfnamefont{P.~A.} \bibnamefont{Souder}},
  \bibnamefont{and} \bibinfo{author}{\bibfnamefont{R.}~\bibnamefont{Michaels}},
  \bibinfo{journal}{Phys. Rev.} \textbf{\bibinfo{volume}{C63}},
  \bibinfo{pages}{025501} (\bibinfo{year}{2001}), \eprint{nucl-th/9912038}.

\bibitem[{\citenamefont{Akimov et~al.}(2017)}]{Akimov:2017ade}
\bibinfo{author}{\bibfnamefont{D.}~\bibnamefont{Akimov}} \bibnamefont{et~al.}
  (\bibinfo{collaboration}{COHERENT}), \bibinfo{journal}{Science}
  \textbf{\bibinfo{volume}{357}}, \bibinfo{pages}{1123} (\bibinfo{year}{2017}),
  \eprint{1708.01294}.

\bibitem[{\citenamefont{Abrahamyan et~al.}(2012)}]{Abrahamyan:2012gp}
\bibinfo{author}{\bibfnamefont{S.}~\bibnamefont{Abrahamyan}}
  \bibnamefont{et~al.}, \bibinfo{journal}{Phys. Rev. Lett.}
  \textbf{\bibinfo{volume}{108}}, \bibinfo{pages}{112502}
  (\bibinfo{year}{2012}), \eprint{1201.2568}.

\bibitem[{\citenamefont{Adhikari et~al.}(2021)}]{Adhikari:2021phr}
\bibinfo{author}{\bibfnamefont{D.}~\bibnamefont{Adhikari}} \bibnamefont{et~al.}
  (\bibinfo{collaboration}{PREX}), \bibinfo{journal}{Phys. Rev. Lett.}
  \textbf{\bibinfo{volume}{126}}, \bibinfo{pages}{172502}
  (\bibinfo{year}{2021}), \eprint{2102.10767}.

\bibitem[{\citenamefont{Reed et~al.}(2021)\citenamefont{Reed, Fattoyev,
  Horowitz, and Piekarewicz}}]{Reed:2021nqk}
\bibinfo{author}{\bibfnamefont{B.~T.} \bibnamefont{Reed}},
  \bibinfo{author}{\bibfnamefont{F.~J.} \bibnamefont{Fattoyev}},
  \bibinfo{author}{\bibfnamefont{C.~J.} \bibnamefont{Horowitz}},
  \bibnamefont{and}
  \bibinfo{author}{\bibfnamefont{J.}~\bibnamefont{Piekarewicz}},
  \bibinfo{journal}{Phys. Rev. Lett.} \textbf{\bibinfo{volume}{126}},
  \bibinfo{pages}{172503} (\bibinfo{year}{2021}), \eprint{2101.03193}.

\bibitem[{\citenamefont{Centelles et~al.}(2009)\citenamefont{Centelles,
  Roca-Maza, Vinas, and Warda}}]{Centelles:2008vu}
\bibinfo{author}{\bibfnamefont{M.}~\bibnamefont{Centelles}},
  \bibinfo{author}{\bibfnamefont{X.}~\bibnamefont{Roca-Maza}},
  \bibinfo{author}{\bibfnamefont{X.}~\bibnamefont{Vinas}}, \bibnamefont{and}
  \bibinfo{author}{\bibfnamefont{M.}~\bibnamefont{Warda}},
  \bibinfo{journal}{Phys. Rev. Lett.} \textbf{\bibinfo{volume}{102}},
  \bibinfo{pages}{122502} (\bibinfo{year}{2009}), \eprint{0806.2886}.

\bibitem[{\citenamefont{Yue et~al.}(2021)\citenamefont{Yue, Chen, Zhang, and
  Zhou}}]{Yue:2021yfx}
\bibinfo{author}{\bibfnamefont{T.-G.} \bibnamefont{Yue}},
  \bibinfo{author}{\bibfnamefont{L.-W.} \bibnamefont{Chen}},
  \bibinfo{author}{\bibfnamefont{Z.}~\bibnamefont{Zhang}}, \bibnamefont{and}
  \bibinfo{author}{\bibfnamefont{Y.}~\bibnamefont{Zhou}}
  (\bibinfo{year}{2021}), \eprint{2102.05267}.

\bibitem[{\citenamefont{Li et~al.}(2020)\citenamefont{Li, Xu, Zhou, Wang, Zhao,
  Chen, and Wang}}]{Li:2019kkh}
\bibinfo{author}{\bibfnamefont{H.}~\bibnamefont{Li}},
  \bibinfo{author}{\bibfnamefont{H.-j.} \bibnamefont{Xu}},
  \bibinfo{author}{\bibfnamefont{Y.}~\bibnamefont{Zhou}},
  \bibinfo{author}{\bibfnamefont{X.}~\bibnamefont{Wang}},
  \bibinfo{author}{\bibfnamefont{J.}~\bibnamefont{Zhao}},
  \bibinfo{author}{\bibfnamefont{L.-W.} \bibnamefont{Chen}}, \bibnamefont{and}
  \bibinfo{author}{\bibfnamefont{F.}~\bibnamefont{Wang}},
  \bibinfo{journal}{Phys. Rev. Lett.} \textbf{\bibinfo{volume}{125}},
  \bibinfo{pages}{222301} (\bibinfo{year}{2020}), \eprint{1910.06170}.

\bibitem[{\citenamefont{Li et~al.}(2018)\citenamefont{Li, Xu, Zhao, Lin, Zhang,
  Wang, Shen, and Wang}}]{Li:2018oec}
\bibinfo{author}{\bibfnamefont{H.}~\bibnamefont{Li}},
  \bibinfo{author}{\bibfnamefont{H.-j.} \bibnamefont{Xu}},
  \bibinfo{author}{\bibfnamefont{J.}~\bibnamefont{Zhao}},
  \bibinfo{author}{\bibfnamefont{Z.-W.} \bibnamefont{Lin}},
  \bibinfo{author}{\bibfnamefont{H.}~\bibnamefont{Zhang}},
  \bibinfo{author}{\bibfnamefont{X.}~\bibnamefont{Wang}},
  \bibinfo{author}{\bibfnamefont{C.}~\bibnamefont{Shen}}, \bibnamefont{and}
  \bibinfo{author}{\bibfnamefont{F.}~\bibnamefont{Wang}},
  \bibinfo{journal}{Phys. Rev.} \textbf{\bibinfo{volume}{C98}},
  \bibinfo{pages}{054907} (\bibinfo{year}{2018}), \eprint{1808.06711}.

\bibitem[{\citenamefont{Sjostrand et~al.}(2006)\citenamefont{Sjostrand, Mrenna,
  and Skands}}]{Sjostrand:2006za}
\bibinfo{author}{\bibfnamefont{T.}~\bibnamefont{Sjostrand}},
  \bibinfo{author}{\bibfnamefont{S.}~\bibnamefont{Mrenna}}, \bibnamefont{and}
  \bibinfo{author}{\bibfnamefont{P.~Z.} \bibnamefont{Skands}},
  \bibinfo{journal}{JHEP} \textbf{\bibinfo{volume}{05}}, \bibinfo{pages}{026}
  (\bibinfo{year}{2006}), \eprint{hep-ph/0603175}.

\bibitem[{\citenamefont{Sjostrand et~al.}(2008)\citenamefont{Sjostrand, Mrenna,
  and Skands}}]{Sjostrand:2007gs}
\bibinfo{author}{\bibfnamefont{T.}~\bibnamefont{Sjostrand}},
  \bibinfo{author}{\bibfnamefont{S.}~\bibnamefont{Mrenna}}, \bibnamefont{and}
  \bibinfo{author}{\bibfnamefont{P.~Z.} \bibnamefont{Skands}},
  \bibinfo{journal}{Comput. Phys. Commun.} \textbf{\bibinfo{volume}{178}},
  \bibinfo{pages}{852} (\bibinfo{year}{2008}), \eprint{0710.3820}.

\bibitem[{\citenamefont{Skands et~al.}(2014)\citenamefont{Skands, Carrazza, and
  Rojo}}]{Skands:2014pea}
\bibinfo{author}{\bibfnamefont{P.}~\bibnamefont{Skands}},
  \bibinfo{author}{\bibfnamefont{S.}~\bibnamefont{Carrazza}}, \bibnamefont{and}
  \bibinfo{author}{\bibfnamefont{J.}~\bibnamefont{Rojo}},
  \bibinfo{journal}{Eur. Phys. J. C} \textbf{\bibinfo{volume}{74}},
  \bibinfo{pages}{3024} (\bibinfo{year}{2014}), \eprint{1404.5630}.

\bibitem[{\citenamefont{Abelev et~al.}(2009)}]{Abelev:2008ab}
\bibinfo{author}{\bibfnamefont{B.~I.} \bibnamefont{Abelev}}
  \bibnamefont{et~al.} (\bibinfo{collaboration}{STAR}), \bibinfo{journal}{Phys.
  Rev.} \textbf{\bibinfo{volume}{C79}}, \bibinfo{pages}{034909}
  (\bibinfo{year}{2009}), \eprint{0808.2041}.

\bibitem[{\citenamefont{Wang and Gyulassy}(1991)}]{Wang:1991hta}
\bibinfo{author}{\bibfnamefont{X.-N.} \bibnamefont{Wang}} \bibnamefont{and}
  \bibinfo{author}{\bibfnamefont{M.}~\bibnamefont{Gyulassy}},
  \bibinfo{journal}{Phys. Rev.} \textbf{\bibinfo{volume}{D44}},
  \bibinfo{pages}{3501} (\bibinfo{year}{1991}).

\bibitem[{\citenamefont{Wang}(1997)}]{Wang:1996yf}
\bibinfo{author}{\bibfnamefont{X.-N.} \bibnamefont{Wang}},
  \bibinfo{journal}{Phys. Rept.} \textbf{\bibinfo{volume}{280}},
  \bibinfo{pages}{287} (\bibinfo{year}{1997}), \eprint{hep-ph/9605214}.

\bibitem[{\citenamefont{Bass et~al.}(1998)}]{Bass:1998ca}
\bibinfo{author}{\bibfnamefont{S.~A.} \bibnamefont{Bass}} \bibnamefont{et~al.},
  \bibinfo{journal}{Prog. Part. Nucl. Phys.} \textbf{\bibinfo{volume}{41}},
  \bibinfo{pages}{255} (\bibinfo{year}{1998}), \eprint{nucl-th/9803035}.

\bibitem[{\citenamefont{Bleicher et~al.}(1999)}]{Bleicher:1999xi}
\bibinfo{author}{\bibfnamefont{M.}~\bibnamefont{Bleicher}}
  \bibnamefont{et~al.}, \bibinfo{journal}{J. Phys.}
  \textbf{\bibinfo{volume}{G25}}, \bibinfo{pages}{1859} (\bibinfo{year}{1999}),
  \eprint{hep-ph/9909407}.

\bibitem[{\citenamefont{Xu et~al.}(2018)\citenamefont{Xu, Wang, Li, Zhao, Lin,
  Shen, and Wang}}]{Xu:2017zcn}
\bibinfo{author}{\bibfnamefont{H.-J.} \bibnamefont{Xu}},
  \bibinfo{author}{\bibfnamefont{X.}~\bibnamefont{Wang}},
  \bibinfo{author}{\bibfnamefont{H.}~\bibnamefont{Li}},
  \bibinfo{author}{\bibfnamefont{J.}~\bibnamefont{Zhao}},
  \bibinfo{author}{\bibfnamefont{Z.-W.} \bibnamefont{Lin}},
  \bibinfo{author}{\bibfnamefont{C.}~\bibnamefont{Shen}}, \bibnamefont{and}
  \bibinfo{author}{\bibfnamefont{F.}~\bibnamefont{Wang}},
  \bibinfo{journal}{Phys. Rev. Lett.} \textbf{\bibinfo{volume}{121}},
  \bibinfo{pages}{022301} (\bibinfo{year}{2018}), \eprint{1710.03086}.

\bibitem[{\citenamefont{Chabanat et~al.}(1997)\citenamefont{Chabanat, Meyer,
  Bonche, Schaeffer, and Haensel}}]{Chabanat:1997qh}
\bibinfo{author}{\bibfnamefont{E.}~\bibnamefont{Chabanat}},
  \bibinfo{author}{\bibfnamefont{J.}~\bibnamefont{Meyer}},
  \bibinfo{author}{\bibfnamefont{P.}~\bibnamefont{Bonche}},
  \bibinfo{author}{\bibfnamefont{R.}~\bibnamefont{Schaeffer}},
  \bibnamefont{and} \bibinfo{author}{\bibfnamefont{P.}~\bibnamefont{Haensel}},
  \bibinfo{journal}{Nucl. Phys.} \textbf{\bibinfo{volume}{A627}},
  \bibinfo{pages}{710} (\bibinfo{year}{1997}).

\bibitem[{\citenamefont{Chabanat et~al.}(1998)\citenamefont{Chabanat, Bonche,
  Haensel, Meyer, and Schaeffer}}]{Chabanat:1997un}
\bibinfo{author}{\bibfnamefont{E.}~\bibnamefont{Chabanat}},
  \bibinfo{author}{\bibfnamefont{P.}~\bibnamefont{Bonche}},
  \bibinfo{author}{\bibfnamefont{P.}~\bibnamefont{Haensel}},
  \bibinfo{author}{\bibfnamefont{J.}~\bibnamefont{Meyer}}, \bibnamefont{and}
  \bibinfo{author}{\bibfnamefont{R.}~\bibnamefont{Schaeffer}},
  \bibinfo{journal}{Nucl. Phys.} \textbf{\bibinfo{volume}{A635}},
  \bibinfo{pages}{231} (\bibinfo{year}{1998}), \bibinfo{note}{[Erratum: Nucl.
  Phys. A643,441(1998)]}.

\bibitem[{\citenamefont{Wang et~al.}(2016)\citenamefont{Wang, Friar, and
  Hayes}}]{Wang:2016rqh}
\bibinfo{author}{\bibfnamefont{X.~B.} \bibnamefont{Wang}},
  \bibinfo{author}{\bibfnamefont{J.~L.} \bibnamefont{Friar}}, \bibnamefont{and}
  \bibinfo{author}{\bibfnamefont{A.~C.} \bibnamefont{Hayes}},
  \bibinfo{journal}{Phys. Rev.} \textbf{\bibinfo{volume}{C94}},
  \bibinfo{pages}{034314} (\bibinfo{year}{2016}), \eprint{1607.02149}.

\bibitem[{\citenamefont{Chamel et~al.}(2009)\citenamefont{Chamel, Goriely, and
  Pearson}}]{Chamel:2009yx}
\bibinfo{author}{\bibfnamefont{N.}~\bibnamefont{Chamel}},
  \bibinfo{author}{\bibfnamefont{S.}~\bibnamefont{Goriely}}, \bibnamefont{and}
  \bibinfo{author}{\bibfnamefont{J.~M.} \bibnamefont{Pearson}},
  \bibinfo{journal}{Phys. Rev.} \textbf{\bibinfo{volume}{C80}},
  \bibinfo{pages}{065804} (\bibinfo{year}{2009}), \eprint{0911.3346}.

\bibitem[{\citenamefont{Zhang and Chen}(2016)}]{Zhang:2015vaa}
\bibinfo{author}{\bibfnamefont{Z.}~\bibnamefont{Zhang}} \bibnamefont{and}
  \bibinfo{author}{\bibfnamefont{L.-W.} \bibnamefont{Chen}},
  \bibinfo{journal}{Phys. Rev.} \textbf{\bibinfo{volume}{C94}},
  \bibinfo{pages}{064326} (\bibinfo{year}{2016}), \eprint{1510.06459}.

\bibitem[{\citenamefont{Zhang and Chen}(2014)}]{Zhang:2014yfa}
\bibinfo{author}{\bibfnamefont{Z.}~\bibnamefont{Zhang}} \bibnamefont{and}
  \bibinfo{author}{\bibfnamefont{L.-W.} \bibnamefont{Chen}},
  \bibinfo{journal}{Phys. Rev.} \textbf{\bibinfo{volume}{C90}},
  \bibinfo{pages}{064317} (\bibinfo{year}{2014}), \eprint{1407.8054}.

\bibitem[{\citenamefont{Moreland et~al.}(2015)\citenamefont{Moreland, Bernhard,
  and Bass}}]{Moreland:2014oya}
\bibinfo{author}{\bibfnamefont{J.~S.} \bibnamefont{Moreland}},
  \bibinfo{author}{\bibfnamefont{J.~E.} \bibnamefont{Bernhard}},
  \bibnamefont{and} \bibinfo{author}{\bibfnamefont{S.~A.} \bibnamefont{Bass}},
  \bibinfo{journal}{Phys.Rev.} \textbf{\bibinfo{volume}{C92}},
  \bibinfo{pages}{011901} (\bibinfo{year}{2015}), \eprint{1412.4708}.

\bibitem[{\citenamefont{Bernhard et~al.}(2016)\citenamefont{Bernhard, Moreland,
  Bass, Liu, and Heinz}}]{Bernhard:2016tnd}
\bibinfo{author}{\bibfnamefont{J.~E.} \bibnamefont{Bernhard}},
  \bibinfo{author}{\bibfnamefont{J.~S.} \bibnamefont{Moreland}},
  \bibinfo{author}{\bibfnamefont{S.~A.} \bibnamefont{Bass}},
  \bibinfo{author}{\bibfnamefont{J.}~\bibnamefont{Liu}}, \bibnamefont{and}
  \bibinfo{author}{\bibfnamefont{U.}~\bibnamefont{Heinz}},
  \bibinfo{journal}{Phys. Rev.} \textbf{\bibinfo{volume}{C94}},
  \bibinfo{pages}{024907} (\bibinfo{year}{2016}), \eprint{1605.03954}.

\bibitem[{\citenamefont{Deng et~al.}(2016)\citenamefont{Deng, Huang, Ma, and
  Wang}}]{Deng:2016knn}
\bibinfo{author}{\bibfnamefont{W.-T.} \bibnamefont{Deng}},
  \bibinfo{author}{\bibfnamefont{X.-G.} \bibnamefont{Huang}},
  \bibinfo{author}{\bibfnamefont{G.-L.} \bibnamefont{Ma}}, \bibnamefont{and}
  \bibinfo{author}{\bibfnamefont{G.}~\bibnamefont{Wang}},
  \bibinfo{journal}{Phys. Rev.} \textbf{\bibinfo{volume}{C94}},
  \bibinfo{pages}{041901} (\bibinfo{year}{2016}), \eprint{1607.04697}.

\bibitem[{\citenamefont{Xu et~al.}(2021)\citenamefont{Xu, Li, Wang, Shen, and
  Wang}}]{Xu:2021vpn}
\bibinfo{author}{\bibfnamefont{H.-j.} \bibnamefont{Xu}},
  \bibinfo{author}{\bibfnamefont{H.}~\bibnamefont{Li}},
  \bibinfo{author}{\bibfnamefont{X.}~\bibnamefont{Wang}},
  \bibinfo{author}{\bibfnamefont{C.}~\bibnamefont{Shen}}, \bibnamefont{and}
  \bibinfo{author}{\bibfnamefont{F.}~\bibnamefont{Wang}},
  \bibinfo{journal}{Phys. Lett. B} \textbf{\bibinfo{volume}{819}},
  \bibinfo{pages}{136453} (\bibinfo{year}{2021}), \eprint{2103.05595}.

\bibitem[{\citenamefont{Adamczyk et~al.}(2014)}]{Adamczyk:2013dal}
\bibinfo{author}{\bibfnamefont{L.}~\bibnamefont{Adamczyk}} \bibnamefont{et~al.}
  (\bibinfo{collaboration}{STAR}), \bibinfo{journal}{Phys. Rev. Lett.}
  \textbf{\bibinfo{volume}{112}}, \bibinfo{pages}{032302}
  (\bibinfo{year}{2014}), \eprint{1309.5681}.

\bibitem[{\citenamefont{Xu et~al.}(2016)\citenamefont{Xu, Yu, Liu, and
  Luo}}]{Xu:2016qjd}
\bibinfo{author}{\bibfnamefont{J.}~\bibnamefont{Xu}},
  \bibinfo{author}{\bibfnamefont{S.}~\bibnamefont{Yu}},
  \bibinfo{author}{\bibfnamefont{F.}~\bibnamefont{Liu}}, \bibnamefont{and}
  \bibinfo{author}{\bibfnamefont{X.}~\bibnamefont{Luo}},
  \bibinfo{journal}{Phys. Rev. C} \textbf{\bibinfo{volume}{94}},
  \bibinfo{pages}{024901} (\bibinfo{year}{2016}), \eprint{1606.03900}.

\bibitem[{\citenamefont{Koch et~al.}(2017)\citenamefont{Koch, Schlichting,
  Skokov, Sorensen, Thomas, Voloshin, Wang, and Yee}}]{Skokov:2016yrj}
\bibinfo{author}{\bibfnamefont{V.}~\bibnamefont{Koch}},
  \bibinfo{author}{\bibfnamefont{S.}~\bibnamefont{Schlichting}},
  \bibinfo{author}{\bibfnamefont{V.}~\bibnamefont{Skokov}},
  \bibinfo{author}{\bibfnamefont{P.}~\bibnamefont{Sorensen}},
  \bibinfo{author}{\bibfnamefont{J.}~\bibnamefont{Thomas}},
  \bibinfo{author}{\bibfnamefont{S.}~\bibnamefont{Voloshin}},
  \bibinfo{author}{\bibfnamefont{G.}~\bibnamefont{Wang}}, \bibnamefont{and}
  \bibinfo{author}{\bibfnamefont{H.-U.} \bibnamefont{Yee}},
  \bibinfo{journal}{Chin. Phys. C} \textbf{\bibinfo{volume}{41}},
  \bibinfo{pages}{072001} (\bibinfo{year}{2017}), \eprint{1608.00982}.

\bibitem[{\citenamefont{Li}(2002)}]{Li:2002qx}
\bibinfo{author}{\bibfnamefont{B.-A.} \bibnamefont{Li}},
  \bibinfo{journal}{Phys. Rev. Lett.} \textbf{\bibinfo{volume}{88}},
  \bibinfo{pages}{192701} (\bibinfo{year}{2002}), \eprint{nucl-th/0205002}.

\bibitem[{\citenamefont{Zhang et~al.}(2018)\citenamefont{Zhang, Zhou, Zhang,
  Zhang, Li, Shao, Sun, and Tang}}]{Zhang:2018hcp}
\bibinfo{author}{\bibfnamefont{S.}~\bibnamefont{Zhang}},
  \bibinfo{author}{\bibfnamefont{L.}~\bibnamefont{Zhou}},
  \bibinfo{author}{\bibfnamefont{Y.}~\bibnamefont{Zhang}},
  \bibinfo{author}{\bibfnamefont{M.}~\bibnamefont{Zhang}},
  \bibinfo{author}{\bibfnamefont{C.}~\bibnamefont{Li}},
  \bibinfo{author}{\bibfnamefont{M.}~\bibnamefont{Shao}},
  \bibinfo{author}{\bibfnamefont{Y.}~\bibnamefont{Sun}}, \bibnamefont{and}
  \bibinfo{author}{\bibfnamefont{Z.}~\bibnamefont{Tang}},
  \bibinfo{journal}{Nucl. Sci. Tech.} \textbf{\bibinfo{volume}{29}},
  \bibinfo{pages}{136} (\bibinfo{year}{2018}), \eprint{1803.05767}.

\end{thebibliography}
\end{document}